# Service Level Agreement for the QoS Guaranteed Mobile IPTV Services over Mobile WiMAX Networks

Mostafa Zaman Chowdhury[1], Bui Minh Trung[1], Yeong Min Jang[1], Young-Il Kim[2], and Won Ryu[2]
[1]Department of Electronics Engineering, Kookmin University, Seoul, Korea
[2]Electronic and Telecommunications Research Institute (ETRI), Korea
E-mail: yjang@kookmin.ac.kr

*Abstract* — **While mobile IPTV services are supported through the mobile WiMAX networks, there must need some guaranteed bandwidth for the IPTV services especially if IPTV and non-IPTV services are simultaneously supported by the mobile WiMAX networks. The quality of an IPTV service definitely depends on the allocated bandwidth for that channel. However, due to the high quality IPTV services and to support of huge non-IPTV traffic over mobile WiMAX networks, it is not possible to guarantee the sufficient amount of the limited mobile WiMAX bandwidth for the mobile IPTV services every time. A Service Level Agreement (SLA) between the mobile IPTV service provider and mobile WiMAX network operator to reserve sufficient bandwidth for the IPTV calls can increase the satisfaction level of the mobile IPTV users. In this paper, we propose a SLA negotiation procedure for mobile IPTV users over mobile WiMAX networks. The Bandwidth Broker controls the allocated bandwidth for IPTV and non-IPTV users. The proposed dynamically reserved bandwidth for the IPTV services increases the IPTV user's satisfaction level. The simulation results state that, our proposed scheme is able to provide better user satisfaction level for the IPTV users.**

## I. Introduction

Internet Protocol Television (IPTV) is gaining recognition as a viable alternative for the delivery of video by telecommunications and cable companies [1]. The mobile IPTV technology enables users to transmit and receive multimedia traffic including television signal, video, audio, text and graphic services through IP-based wireless networks with full support of Quality of Service (QoS) and Quality of Experience (QoE), security, mobility, and interactive functions [2]. In case of mobile IPTV, users can enjoy IPTV services anywhere and even while on the move using the wireless access. IPTV uses standard networking protocols. So, it promises lower costs for operators and lower prices for users. The mobile WiMAX networks will be the key networks in supporting the promising and extremely demanded IPTV services with full mobility support. In general, mobile IPTV services can be classified into on-demand content, live content, managed services, and unmanaged services.

The mobile IPTV services can be provided using several access networks. The mobile IPTV user may connect with the existing macrocellular networks or mobile WiMAX networks or femtocell networks or others wireless networks. Using different wireless access networks, the IPTV services can be provided at anytime and anywhere. However, the IPTV over mobile WiMAX networks will be an effective but challenging for the several reasons. A mobile WiMAX network may use exclusively for the mobile IPTV purposes or IPTV services may jointly served with non-IPTV services (e.g., internet, smart phone, and etc.) by the mobile WiMAX network. The advantages of mobile WiMAX networks to support the mobile IPTV services will be discussed in the later part of this paper.

Even though there are many advantages to use the IP based mobile WiMAX networks, there are also some challenges. Due to the high quality IPTV services and to support of huge non-IPTV traffic over mobile WiMAX networks, it is not possible to guarantee the sufficient amount of the limited mobile WiMAX bandwidth for the mobile IPTV services every time. A Service Level Agreement (SLA) [3]-[5] between the mobile IPTV service provider and mobile WiMAX network operator to reserve sufficient bandwidth for the IPTV calls can increase the satisfaction level of the mobile IPTV users. For the inequality between mobile WiMAX network capacity and bandwidth demanded by non-IPTV services and IPTV services, some requested mobile IPTV calls are blocked and some ongoing IPTV calls are dropped or quality degraded. Until now, no researcher proposed the SLA architecture for the mobile IPTV services over mobile WiMAX networks. Due to our SLA negotiation, mobile WiMAX network reserve sufficient bandwidth for IPTV calls. The Bandwidth Broker can be used to efficiently distribute the bandwidth for mobile IPTV users and non-IPTV users.

This paper is organized as follows. Section II provides study about the necessity of SLA for the mobile IPTV deployment over mobile WIMAX networks. The proposed SLA architecture is presented in Section III. In Section IV, we presented the bandwidth reservation scheme for the proposed SLA framework. The simulation results for the proposed scheme are performed in Section V. Finally, we concluded our paper in Section VI.

## II. Why SLA for the Mobile IPTV Deployment over Mobile WiMAX Networks?

Recently mobile WiMAX has become very popular for its capability to support high data rates in cellular environments and QoS for different applications. Mobile

WiMAX can effectively support video streaming, VoIP, and data services. The mobile WiMAX is an all IP based access network. It offers the transparency for the packet based core networks. The mobile WiMAX radios are designed not to add any harm to the users. The mobile WiMAX systems are suitable for the delivery of IP based quadruple play services; broadband internet access, television, and telephone with wireless service provisions. These and some additional features of mobile WiMAX networks make mobile WiMAX a best choice for the mobile IPTV services rather than cable, DSL, cellular and other networks.

Mobile WiMAX networks offer the wireless connectivity in the access networks that can support IPTV services with mobility. The standards of the IEEE 802.16 family [6] provide fixed and mobile broadband wireless access (BWA). IEEE 802.16 family is a promising technology to deliver multiple high-data-rate services over large areas. Based on complexity and flexibility management of the medium access control (MAC) and physical (PHY) layers, the IEEE 802.16 family is expected to support QoS [7]. IEEE 802.16m can support advanced air interface with data rates of up to 100 Mbit/s for the mobile users and up to 1 Gbit/s for the fixed users. IEEE 802.16 can also provide secure delivery of content and support mobile users at vehicular speeds. The bandwidth of IEEE 802.16 is scalable and manageable. The contents are encrypted for secured transmission that is very important for the IPTV services. The installation and maintenance costs of mobile WiMAX systems are also at a fraction of the costs of wired access networks. Considering the advantages of the features of mobile WiMAX systems, IPTV services can be designed, delivered, and managed cost effectively without compromising the video and audio quality [1]. WiMAX MAC layer can support real time poling services that ensure required bandwidth and minimum latencies for the IPTV video services through QoS. It uses OFDM and OFDMA PHY layers. Hence, the OFDM and OFDMA PHY layers are resilient to multipath fading channels. Moreover, it uses adaptive modulation schemes and forward error correction (FEC) to increase service quality and also since WiMAX PHY supports varying frame sizes and scalable bandwidth, WiMAX is an ideal choice for IPTV applications [1].

Even though there are many advantages to using IP based mobile WiMAX networks for mobile IPTV deployment, there are also some challenges. Transmitting IPTV services over IP faces significant QoS challenges for the service provider. The IPTV services require higher bandwidth compared to the other applications such as voice and data. IPTV has a typical bandwidth requirement of 2 Mbps/channel to 8 Mbps/channel (HDTV) and it requires a broadband connection with QoS support [8]. Hence, the guarantee of bandwidth per broadcasting channel is very important whenever the mobile IPTV services are supported by mobile WiMAX networks especially if the non-IPTV and IPTV services are jointly supported by the mobile WiMAX networks.

The mobile IPTV service provider and mobile WiMAX network operator may not be the same. Hence, different operators are involved for the IPTV services over mobile WiMAX networks. Due to these heterogeneous operators, there should have inter-operator agreements to provide end-to-end QoS for the IPTV users. Quality of the backhaul connection typically depends on mobile WiMAX capacity, overall backhaul network load, and bandwidth management. Mobile IPTV traffic competes with other multiple non-IPTV traffic to send data through the mobile WiMAX network. This may create a traffic bottleneck due to the lack of backhaul capacity and unavailability of a prioritization scheme. Thus, the quality of IPTV services may degrade significantly. Hence, no prioritization on backhaul link can degrade the IPTV services. Thus, to guarantee the QoS of IPTV services, there should be SLA and tight inter-operability between the IPTV service provider and mobile WiMAX network provider. A SLA between IPTV service provider and mobile WiMAX network provider can be developed so both operators benefit and the QoS level of mobile IPTV users is ensured. The author in [3] explains that a typical SLA contains basic information about the service, a description of the nature of service to be provided, components, the expected performance level of the service (specifically its reliability and responsiveness), the procedure for reporting problems with the service, the time frame for response and problem resolution, the process for monitoring and reporting the service level, and the consequences for the service provider not meeting its obligations.

### III. Proposed SLA Architecture

The availability of requested bandwidth by mobile IPTV users is the key element for ensuring the QoS. The same mobile WiMAX network is used for several non-IPTV services simultaneously with the IPTV services. So, for the heavy non-IPTV traffic conditions, sufficient bandwidth to support mobile IPTV users cannot be ensured. The allocated bandwidth for the mobile IPTV traffic can be increased by degrading the allocated bandwidth for the non-IPTV services. To ensure the sufficient bandwidth for the mobile IPTV users, a SLA negotiation between the IPTV service provider and mobile WiMAX network provider is needed.

#### A. Proposed Network Architecture

Figure 1 shows our proposed SLA network architecture for the mobile IPTV deployment over mobile WiMAX networks. The video server in the content provider part, store audio/video (A/V) content which are encoded and compressed from live and pre-recorded programs. Video servers are either centralized or distributed in core networks. The IPTV service provider is responsible for the type of service that will be provided to users. There is a service level agreement between the IPTV service provider and the mobile WiMAX network operator. The bandwidth broker (BB) executes the SLA between the IPTV service provider and the mobile WiMAX network operator. The Bandwidth BB efficiently distributes the bandwidth among the IPTV services and non-IPTV services.

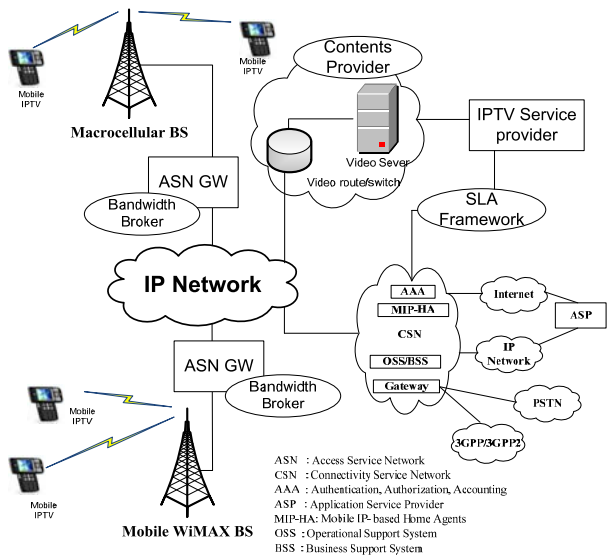

**Figure 1.** Network architecture for the SLA based mobile IPTV services over mobile WiMAX networks

### B. Proposed Framework of the SLA Execution

A SLA negotiation between the mobile WiMAX network operator and mobile IPTV service provider can provide sufficient bandwidth for mobile IPTV users. For the inequality between mobile WiMAX Network capacity and bandwidth demanded by non- IPTV services and IPTV services, some requested mobile IPTV calls are blocked and some ongoing IPTV calls are dropped or quality degraded. Due to the SLA negotiation, mobile WiMAX network reserve sufficient bandwidth for IPTV calls. The bandwidth broker [9] is used to allocate bandwidth for mobile IPTV users and non-IPTV users. The BB makes a policy access, resource reservation and admission control decisions for the SLA framework. From the previous history of requested bandwidth by mobile IPTV calls, Bandwidth Broker reserves bandwidth for mobile IPTV calls. The BB calculates the average requested bandwidth from the previous call history for a certain period of time. Then the BB allocates this amount of bandwidth for mobile IPTV users from the mobile WiMAX network capacity. The remaining bandwidth can be used by other non-IPTV services. BB considers only a certain period of recent call history, therefore, there is not much more difference between the reserve bandwidth and the requested bandwidth by mobile IPTV users. Thus the bandwidth utilization of mobile WiMAX network is not decreased but satisfaction level of mobile IPTV user's is increased significantly.

The BB is an agent responsible for allocating preferred service to users as requested. BB has a policy database to keep the information on who can do what, when and a method of using that database to authenticate requesters. When an allocation is desired for a particular flow, a request is sent to the BB. The requests include the service type, target rate, maximum burst, and the time period when service is required. The BB verifies the unallocated bandwidth whether sufficient to meet the request or not.

Under the BB architecture, admission control, resource provisioning, and other policy decisions are performed by a centralized BB in each network domain [10], BB may handle the mobile WiMAX network bandwidth for mobile IPTV users efficiently.

Figure 2 shows the framework for our proposed scheme. There is a SLA negotiation between the mobile WiMAX network operator, IPTV content provider, and the mobile IPTV service provider to ensure sufficient bandwidth for the mobile IPTV users. The BB is the central logical entity that is responsible for the execution of SLA negotiation and resource allocation. The BB knows the current and previous states of bandwidth utilization information. Using this information, BB calculates the reserve bandwidth. The BB monitors the services and their bandwidth allocation, requested bandwidth, type of services, and etc. of the client networks (mobile WiMAX network operator, IPTV content provider and mobile IPTV service provider). The mobile WiMAX network installs a monitoring scheme to measure the bandwidth allocation among different access networks. The Database stores the information of client networks, SLA policy, and call history. The BB can configure the policy of bandwidth distribution according to the feedback from the monitoring and external input policy. The amount of reserve bandwidth for mobile IPTV users is calculated according to the configuration policy, monitoring result, and collected information from Database. The distribution of bandwidth can be controlled by a rate control device that provides the ability to reserve different bandwidth for different applications. Then the bandwidth is allocated among different services based on SLA.

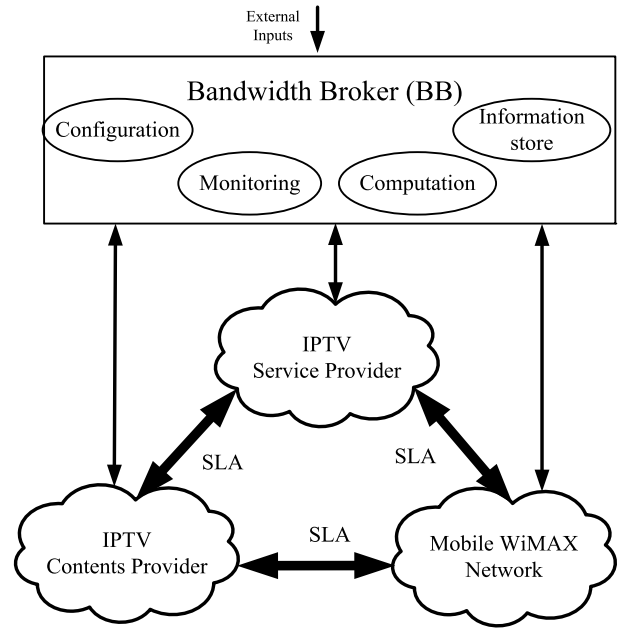

**Figure 2.** SLA between the mobile WiMAX networks and other networks and service provider to guarantee the sufficient bandwidth for the IPTV traffic

## IV. Proposed Bandwidth Reservation Scheme to Support QoS for IPTV Users

A channel will be broadcasted by the mobile WiMAX network only when at least one user requests to watch that channel. The bandwidth satisfaction level of the requested IPTV users depends on the availability of bandwidth for a requested channel. Available bandwidth equal to or greater than requested bandwidth promotes the highest satisfaction levels. It means that there is no bandwidth degradation of active broadcasting channels to accept a new channel request. When the available bandwidth is less than the requested bandwidth by a new channel, there is the possibility of some quality degradation of ongoing broadcasting channels or dropping of some ongoing broadcasting channels or blocking of new channel request. Hence, we define a term, "satisfaction level," that depends on the availability of bandwidth for the new channel request or allocated bandwidth for the active channels. This scheme calculates the amount of reserve bandwidth required for IPTV channels. The SLA between IPTV service provider and mobile WiMAX network operator ensure the bandwidth reservation for the IPTV channels. The BB manages and monitors the distribution of bandwidth among IPTV services and non-IPTV services over mobile WiMAX.

Suppose total capacity of mobile WiMAX networks, required bandwidth by existing non-IPTV services, and bandwidth demand by the IPTV services are denoted by $C$, $B_I(t)$, and $B_{IPTV}(t)$, respectively. Then, the satisfaction level ($SL$) of IPTV calls is calculated as follows.
Available bandwidth for IPTV calls is:

$$B_A(t) = C - B_I(t) \tag{1}$$

if $B_A(t) \geq B_{IPTV}(t)$, then the satisfaction level is:

$$SL(t) = 1 \tag{2}$$

if $B_A(t) < B_{IPTV}(t)$, then the satisfaction level is:

$$SL(t) = \frac{B_A(t)}{B_{IPTV}(t)} \tag{3}$$

Let, $T$ be the time period for which every $t_1$ interval the required bandwidth for the IPTV services is observed. Then, the required dynamically reserved bandwidth $B_R(t)$ for the IPTV calls in proposed scheme can be calculated as:

$$B_R(t) = min\left(\frac{\sum_{n=1}^{N} B_{IPTV}(t - nt_1)}{N}, \Gamma\right) \tag{4}$$

where $N = \frac{T}{t_1}$ is the total number of samples for the calculation of reserved bandwidth at time $t$. $B_{IPTV}(t - nt_1)$ is the total required bandwidth for the IPTV services at $nt_1$ time past form the present time. $\Gamma$ is the maximum allowable bandwidth for the IPTV services.

If $B_R(t) > B_A(t)$, then the borrowing bandwidth $B_B(t)$ from the existing non-IPTV calls is:

$$B_B(t) = B_R(t) - B_A(t) \tag{5}$$

The allocated bandwidth for each channel of IPTV services in the non-SLA scheme, where all the IPTV calls and non-IPTV calls are degraded with equal rate can be expressed as:

$$\beta_{IPTV} = \begin{cases} \beta_{max}, & \beta_{max} N_{IPTV} + B_I(t) \leq C \\ \frac{C}{\beta_{max} N_{IPTV} + B_I(t)} \beta_{max}, & \beta_{max} N_{IPTV} + B_I(t) > C \end{cases} \tag{6}$$

where $\beta_{IPTV}$ is the allocated bandwidth for each IPTV channel, $N_{IPTV}$ is the existing number of active IPTV channels. $\beta_{max}$ is the maximum required bandwidth for each IPTV channel.

Now, the allocated bandwidth for each channel of IPTV services in the proposed SLA scheme is:

$$\beta'_{IPTV} = \begin{cases} \beta_{max}, & \frac{B_R(t)}{N_{IPTV}} \geq \beta_{max} \\ \frac{B_R(t)}{N_{IPTV}}, & \text{other cases} \end{cases} \tag{7}$$

The SLA between mobile WiMAX operator and mobile IPTV service provider ensures minimum $B_R(t)$ bandwidth for IPTV calls. Hence, available bandwidth for IPTV calls is always greater than or equal to $B_R(t)$.

## V. Simulation Result

The performance of the proposed scheme is performed using simulation results. We performed the satisfaction level of IPTV users that depends on the availability of requested bandwidth. Table 1 shows the basic assumptions and parameters for our simulation. We consider 20 mobile WiMAX networks for our simulation. The results are taken from the average of 20 mobile WiMAX networks cases.

**Table 1. Simulation parameters**

| | |
|---|---|
| Mobile WiMAX network capacity [Mbps] | 60 |
| Non-IPTV traffic arrival | Poisson |
| IPTV call arrival | Poisson |
| Requested BW (maximum required bandwidth) by a IPTV channel [Mbps] | 2 |
| Maximum number of IPTV channels | 30 |
| Minimum required bandwidth to broadcast a IPTV channel [Mbps] | 1 |
| Maximum allowable bandwidth for the IPTV services [Mbps] | 40 |
| $t_1$ [minute] | 1 |
| T [minute] | 60 |

Figure 3 compares the satisfaction level of a IPTV users for the proposed SLA scheme and a non-SLA scheme for various non-IPTV mobile WiMAX traffic condition. We assume average 20 number of active IPTV channels for this comparison. It shows that, the proposed scheme can provide very high user satisfaction level i.e. very high quality IPTV services even for very high non-IPTV traffic condition. Hence, our scheme reserve a dynamic amount of bandwidh for the active and incoming IPTV traffic calls. Normally any bandwidth reservation scheme reduces the overall bandwidth utilization of the system. However, the proposed scheme does not reduce the overal mobile WiMAX bandwidth utilization. Figure 4 shows that the bandwidth utilization for both the SLA based proposed scheme and non-SLA scheme are almost same. Hence, the IPTV users can enjoy better quality of services even in heavy non-IPTV traffic condition.

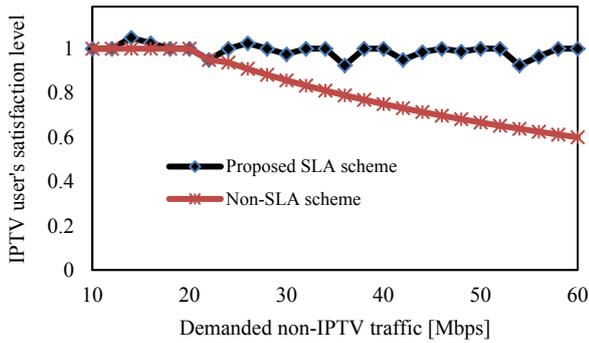

**Figure 3.** Comparison of IPTV user's satisfaction level for various non-IPTV traffic condition

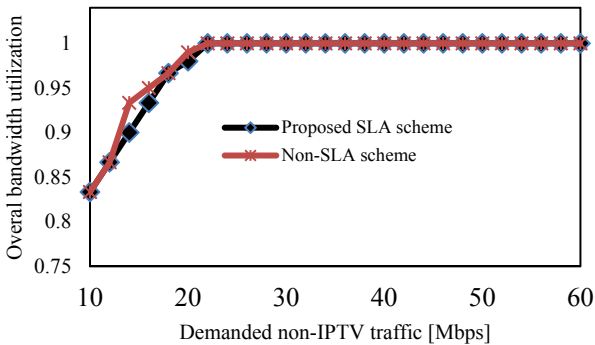

**Figure 4.** Comparison of bandwidth utilization for the proposed SLA scheme and the non-SLA scheme

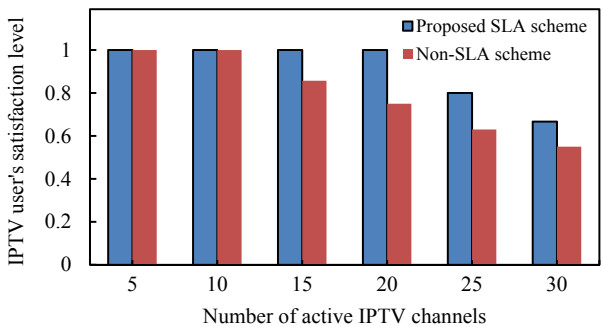

**Figure 5.** Comparison of IPTV user's satisfaction level for various IPTV traffic condition

Figure 5 shows the IPTV user's satisfaction level comparison for the various IPTV traffic condition. It shows that the user satisfaction level i.e. IPTV channel condition only degraded if the number of active IPTV channel is very high. This is happen because of the allowable threshold bandwidth for the IPTV channels. Hence, the above results in Figs. 3, 4, and 5 clearly show the benefit of the proposed scheme. So, it can be concluded from the results, the proposed scheme is very much effective for mobile IPTV deployment over the mobile WiMAX network.

## VI. Conclusion

To support the IPTV services over the wireless link, mobile WiMAX network is an excellent approach for its capability to support high data rate in cellular environments and QoS for different applications. However, the provision of mobile IPTV services using mobile WiMAX networks will be successful if sufficient amount of resources can be guaranteed for the mobile IPTV users. The SLA between the mobile WiMAX network operator and mobile IPTV service provider ensure the sufficient resources for the mobile IPTV users especially when the same mobile WiMAX network simultaneous provides IPTV services and non-IPTV services.

From the previous call history, our proposed scheme calculates the approximate bandwidth for upcoming and active mobile IPTV calls. This amount of bandwidth is reserved for IPTV calls. The simulation results show that our scheme can effectively increase the IPTV users' satisfaction levels without significantly decreasing the bandwidth utilization of the overall system. Thus, our approach will be a promising bandwidth allocation technique to ensure the QoS for the IPTV services over mobile WiMAX networks.

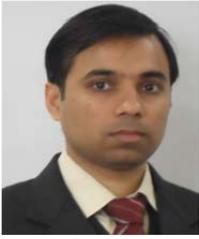
**Mostafa Zaman Chowdhury** received his B.Sc. in Electrical and Electronic Engineering (EEE) from Khulna University of Engineering and Technology (KUET), Bangladesh in 2002. In 2003, he joined the EEE department of KUET as a faculty member. He received his M.Sc. in Electronics Engineering from Kookmin University, Korea in 2008. Currently he is working towards his Ph.D. degree in the department of Electronics Engineering at the Kookmin University. His research interests include convergence networks, QoS provisioning, mobile IPTV, femtocell networks, and VLC networks.

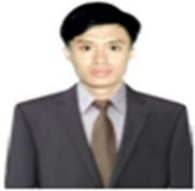
**Bui Minh Trung** received B.Sc. in Telecommunication from Ho Chi Minh city University of Technology (HCMUT), Vietnam in 2010. Currently he is continuing his Master degree in Kookmin University. His current research interests focus on convergence networks and mobile IPTV.

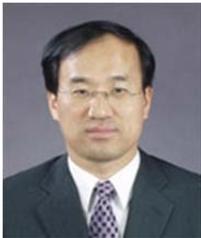
**Yeong Min Jang** received the B.E. and M.E. degree in Electronics Engineering from Kyungpook National University, Korea, in 1985 and 1987, respectively. He received the PhD degree in Computer Science from the University of Massachusetts, USA, in 1999. He worked for ETRI between 1987 and 2000. Since Sept. 2002, he is with the School of Electrical Engineering, Kookmin University, Korea. His research interests are IMT-advanced, radio resource management, and convergence networks.

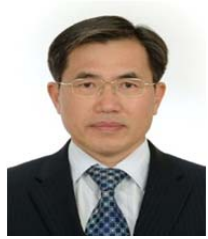
1985년: 경희대학교 전자공학 학사
1988년: 경희대학교 전자공학 석사
1996년: 경희대학교 전자공학 박사
1985년~1986년: 삼성전자
1988년~현재: 한국전자통신연구원 책임연구원, 모바일서비스구조연구팀
1994년: 정보통신기술사, 1995년: 전기통신기술사
2007년: MUST(Mongolian University of Science and Technology) 명예교수
2007년: 과학기술연합대학원대학교(UST)겸임교수
2008년: 한밭대학교 겸임교수, 북경교통대(Beijing Jiaotong University) 고문교수

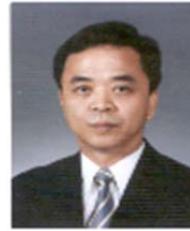
**Won Ryu** received the BS degree in computer science and statistics from Pusan National University, Busan, Korea, in 1983, and the MS degree in computer science and statistics from Seoul National University, Seoul, Korea, in 1988. He received his PhD degree in information engineering from Sungkyunkwan University, Kyonggi, South Korea, in 2000. Since 1989, he has been a managing director with the Smart screen convergence research department, ETRI, Daejeon, Korea. Currently, his research interests are IPTV, Smart TV, IMT-advanced, and convergence services and networks.